\documentclass[epj]{svjour}
\usepackage{graphics}
\begin{document}
\newcommand{\alphab} {{\overline \alpha_s}}
\newcommand{\xp}     {\ensuremath{x_p}}
\newcommand{\xmu}     {\ensuremath{x_\mu}}
\title{Studies of QCD at LEPII}
\subtitle{}
\author{Hasko Stenzel\inst{1}
}                     
\institute{II. Physikalisches Institut, JLU Giessen, Heinrich-Buff Ring 16, 
D-39392 Giessen\\ \email{hasko.stenzel@cern.ch}}
\date{\today}
%
\abstract{A combination of measurements of $\alpha_s$ in $e^+e^-$ annihilation at 
LEP is presented. 
Distributions of various event-shape variables measured
by ALEPH, DELPHI, L3 and OPAL at centre-of-mass energies 
from 41 GeV to 206 GeV are analysed in a common theoretical framework 
using $\cal{O}$$(\alpha_s^2)$+NLLA predictions. The dominant theoretical uncertainties 
associated to missing higher orders are studied in detail. 
%
} 
\maketitle
\section{Introduction}
\label{intro}
The most important parameter of QCD, the strong coupling constant $\alpha_s$, 
has been determined at LEP in different processes 
and with various observables over a wide range of energies. Here measurements using
 event-shape variables provided by the four LEP experiments are combined.   
Six observables are analysed, these are thrust, heavy jet mass, wide and total 
jet broadening, C-parameter and the three-jet resolution parameter $-\ln y_3$. 
The data have been collected at the LEPII-energies in the range from 133 GeV 
to 206 GeV, at the Z boson peak and at lower energies from 41 GeV to 85 GeV 
using radiative events. The combination of the measurements from different 
variables, energies and experiments takes into account correlations between 
the measurements. Theoretical uncertainties are re-evaluated for the individual 
input measurements and propagated to the combined values. This combination is performed 
by the LEP QCD working group \cite{qcdwg}.  
\section{Input measurements}
\label{sec:input}
The latest, partially still preliminary measurements of $\alpha_s$ 
provided by ALEPH \cite{inputaleph}, DELPHI \cite{inputdelphi}, L3 \cite{inputl3} 
and  OPAL \cite{inputopal} serve as input data for the combination. Alltogether 
194 input values enter the combination precedure.  
The measurements are consitently extracted from the event-shape distribution 
using a theoretical description recommended by the LEP QCD working group \cite{qcdwg}. 
Statistical and experimental systematic uncertainties are evaluated by the experiments. 
In order to study hadronisation uncertainties the experiments have provided measurements 
using three generators PYTHIA, HERWIG and ARIADNE for hadronisation corrections. 
The dominant uncertainty 
is perturbative, and it is highly correlated between the input measurements. A special 
treatment is applied to calculate the theoretical uncertainty for each input 
measurement, as explained in the next section.    
\section{Theoretical predictions and uncertainties}
\label{sec:theo}
\subsection{Predictions}
To second order in $\alpha_s$, the
distribution of a generic event-shape variable $y\, (y=1-T, \rho,
B_W, B_T, y_3, C)$  is given by:
 \begin{eqnarray*}\label{fixed}
\frac{1}{\sigma_{tot}}\; \frac{d\sigma(y)}{dy}&=&
\alphab(\mu^{2})A(y)+\left(\alphab(\mu^{2})\right)^{2} \times \\ 
& & \left[A(y) 2\pi b_0 \ln\left(\frac{\mu^{2}}{s}\right)+B(y)\right] \; ,
\end{eqnarray*}
where $\alphab=\frac{\alpha_s}{2\pi}$ and $b_0=\frac{33-2n_{f}}{12\pi}$. 
The resummed calculations are applied to the cumulative cross section 
\begin{displaymath}
R(y,\alpha_s) \equiv \frac{1}{\sigma_{tot}}\int_{0}^{y}\frac{d\sigma(x,\alpha_s)}{dx}dx \; \; .
\end{displaymath}
The fit function consists of two components, the fixed order term and an expression 
resumming leading and next-to-leading logarithms to all orders in $\alpha_s$. 
The recommended \cite{theory} prescription to merge the two calculations and to subtract 
off double counting terms is the Log(R) matching scheme
\begin{eqnarray*}\label{logR}
\ln R(y,\alpha_s) & = & L g_1(\alpha_s L) +g_2(\alpha_s L) - (G_{11}L +G_{12}L^2)\alphab \\
        &  & \hspace*{-2.cm} -(G_{22}L^2 +G_{23}L^3)\alphab^2
          + {\cal A}(y)\alphab +\left[{\cal B}(y)- \frac{1}{2}{\cal A}^2(y)\right]\alphab^2 \; , 
\end{eqnarray*}
with $L=\ln(y_0/y)$ where $y_0=1$ for $y=1-T, \rho, y_3, B_T, B_W$
and $ y_0=6$ for $C$. An improved version, the modified Log(R) scheme  
is used in the fit functions to determine $\alpha_s$, which vanishes at a given 
phase space limit $y_{max}$. To fulfil this constraint $L$ is replaced by     
\begin{displaymath}
\tilde{L} = \frac{1}{p}\ln\left( \left(\frac{y_0}{y}\right)^p -
\left(\frac{y_0} {y_{max}}\right)^p +1\right)\; ,
\end{displaymath}
with the modification degree power $p=1$. A detailed collection of the 
formulae and numerical values for $y_{max}$ and the matching coefficients are given 
in \cite{theory}.
\subsection{Uncertainties}
Theoretical uncertainties stemming from unknown higher orders in the perturbat\-ion 
series are inherently difficult to assess. The method adopted here includes a 
variation of the renormalisation scale $\xmu$ from $0.5$ to 2 and a new test 
which re-scales the logarithmic variable $L\rightarrow \ln\frac{y_0}{x_Ly}$ in 
the range $2/3 < x_L < 3/2$, as suggested in \cite{theory}. In addition a different 
matching scheme and different implementations of the kinematic modification are tested. 
Different sources of the perturbative uncertainty are combined with the uncertainty 
band method. This method calculates the uncertainty for a given variable and 
experimentally used fit range with a common value of $\alpha_s(M_Z)$, the latter being the 
result of a first combination iteration of all input measurements. The uncertainties 
of the distributions are calculated first, taking the prediction of the modified Log(R) 
matching scheme as reference theory and constructing an uncertainty band from the 
largest deviation with respect to the reference in each bin out of the theoretical 
variants mentioned above. In a second step the reference theory is calculated with 
variable $\alpha_s$ and the smallest resp. largest value yielding a prediction inside 
the uncertainty band defines the theoretical uncertainty of $\alpha_s$. The method is 
illustrated in fig.~\ref{fig:uband} taking C-parameter as example.        
\setlength{\unitlength}{1.0 cm}
\begin{figure}[h]
\begin{center}
\begin{picture}(8.0,8.5)
\put(-0.4,-0.5){\includegraphics{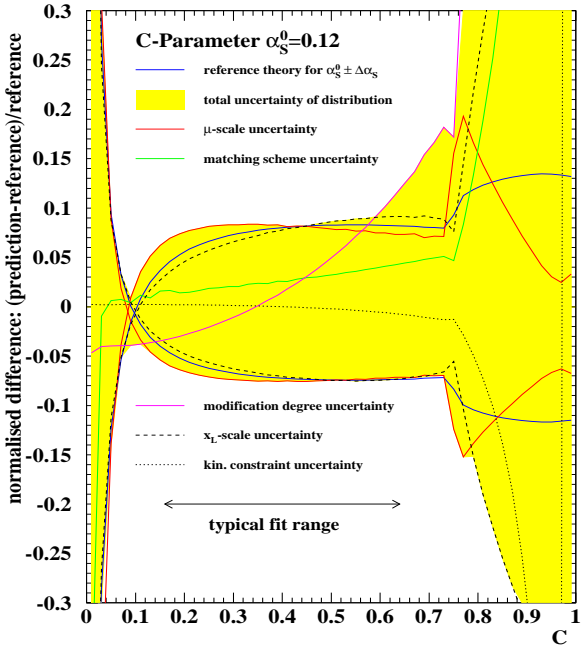}}
\end{picture}
\end{center}
\caption{Theoretical uncertainties for C-parameter. The grey area corresponds
to the total perturbative uncertainty, dotted and dashed lines show
individual contributions. The full lines represent reference predictions with 
values of $\alpha_s$ shifted by its uncertainty.}
\label{fig:uband}
\end{figure}
The main two components to the uncertainty are the variations of the scales $\xmu$ 
and $x_L$, which contribute to a similar amount but in different regions of the 
distributions. 
\section{Combination procedure}
\label{sec:comb}
The procedure to combine the 194 measurements begins with the construction of 
a 194 $\times$ 194 covariance matrix $V$, which relates the uncertainties for all pairs 
of input measurements. The covariance matrix is decomposed into four components
 corresponding to the four sources of errors:
\begin{displaymath}
V_{ij}^\mathrm{total} =
V_{ij}^\mathrm{stat.}\;+\;V_{ij}^\mathrm{exp.}\;+\;
V_{ij}^\mathrm{had.}\;+\;V_{ij}^\mathrm{th.}\;\;\;.
\end{displaymath}
The best combined value $\hat\alpha_s$ yielding the minimal error is a weighted 
average with weights obtained as follows
\begin{displaymath}
\hat\alpha=\sum_i w_i \, \alpha_s^i \;\;,\;\;\mathrm{where}\;\;\;
w_i=\frac{\sum_{j}\;(V^{-1})_{ij}}{\sum_{jk}\;(V^{-1})_{jk}}\;\;\;.
\label{weightsformula}
\end{displaymath}
The uncertainty of the combination is given for each component as function of the 
weights, e.g.  
for the statistical uncertainty     
$\sigma^2_\mathrm{stat.}\!\!=\sum_{ij} w_i\,V_{ij}^\mathrm{stat.}\,w_j $. 
A specific treatment of the correlation of each uncertainty component concerning 
the off-diagonal matrix elements of $V$ is applied.
 
The statistical uncertainties are uncorrelated between different
  experiments and between different energies, but correlated for
  different observables using the same events. These correlations
  have been estimated numerically, using a large number of simulated
  datasets.

The experimental systematic uncertainties are considered to be
  uncorrelated between different experiments. Between different energies and/or different
  observables of the same experiment, the ``minimum overlap'' assumption is made:
$V_{ij}^\mathrm{exp.}=\left[\mathrm{min}(
\sigma^\mathrm{exp.}_i, \sigma^\mathrm{exp.}_j)\right]^2$. 

Studies of the LEP QCD 
working group revealed that the hadronisation uncertainty is largely uncorrelated 
between experiments because of different tunings and versions of the generators. 
Therefore, the off-diagonal matrix elements of $V^\mathrm{had.}$ are set to zero 
and the hadronisation uncertainty is determined by repeating the combination 
procedure for each of the generators separately. The nominal results uses PYTHIA 
for corrections and as uncertainty the standard deviation of the three combinations 
is taken. For the combinations by energy the raw hadronisation uncertainty 
is fitted by a power law form $a+b/Q$ in order to suppress statistical fluctuations. 

The theory uncertainties are expected to be highly correlated
  between all pairs of measurements of the same observable and to
  varying extents between measurements of different observables. The exact determination 
of the correlation, however, turned out to be difficult. Various attempts indicated 
correlation coefficients of the order of 90$\%$, but the result of the combination, 
in particular its theoretical uncertainty, depends strongly on the assumptions made. 
A more reliable approach is chosen, setting again the off-diagonal matrix elements 
of $V^\mathrm{th.}$ to zero and repeating the combination for two additional alternative 
input values of $\alpha_s^0\pm \Delta\alpha_s^\mathrm{th.}$, where $\alpha_s^\mathrm{th.}$ 
is obtained with the uncertainty band method. As perturbative uncertainty 
the difference between the nominal combined result and the alternative combinations is taken. 
\section{Results}
\label{sec:res}
The combination is carried out for different subsets of variables and energies, the 
full results are given in \cite{qcdwg}. The main result is summarised in Table 
\ref{tab:res}, where all variables and experiments are combined for 
different ranges in energy.       
\begin{table}
\caption{Combinations of $\alpha_s(M_Z)$ for different energy ranges.}
\label{tab:res}       
\begin{tabular}{lcccc}
\hline\noalign{\smallskip}
 & $Q<M_Z$ & $Q=M_Z$ & $ Q>M_Z$ & All \\
\noalign{\smallskip}\hline\noalign{\smallskip}
$\alpha_s(M_Z)$ & 0.1208 & 0.1200 & 0.1201 & 0.1201 \\ 
stat. error & $\pm 0.0012$ & $\pm 0.0002$ & $\pm 0.0005$ & $\pm 0.0003$ \\ 
exp. error & $\pm 0.0023$ & $\pm 0.0008$ & $\pm 0.0010$ & $\pm 0.0009$ \\ 
had. error & $\pm 0.0032$ & $\pm 0.0010$ & $\pm 0.0007$ & $\pm 0.0009$ \\ 
th. error & $+ 0.0052$ & $+ 0.0048$ & $+ 0.0044$ & $+ 0.0046$ \\ 
          & $- 0.0050$ & $- 0.0048$ & $- 0.0045$ & $- 0.0047$ \\ 
total error & $\pm 0.0066$ & $\pm 0.0050$ & $\pm 0.0047$ & $\pm 0.0048$ \\ 
\noalign{\smallskip}\hline
\end{tabular}
\end{table}
The best combination with the smallest total uncertainty is obtained with 
the LEPII data alone, combining all data yields a larger uncertainty. This is mainly due 
to the theoretical uncertainty, which is decreasing with increasing energy, scaling 
with $\alpha_s^3$. The hadronisation uncertainty is deceasing as well, scaling with $1/Q$. 

\begin{figure}[h]
\begin{center}
\begin{picture}(8.5,7.)
\put(-2.0,-5.5){\includegraphics{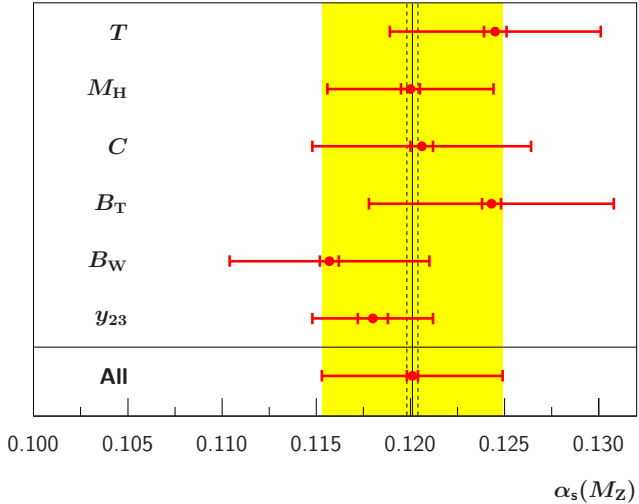}}
\end{picture}
\end{center}
\caption{Combinations of $\alpha_s$ per variable.}
\label{fig:variable}
\end{figure}

The perturbative uncertainties are the key element limiting the precision of present 
elements, and their exact determination is cumbersome. As a cross check, results from 
different variables are compared in fig. \ref{fig:variable}. As each variable receives 
possibly different higher order contribution, the spread of results in $\alpha_s$ indicates 
their size. The RMS between results from different variables is 0.0016 with a maximum 
spread of 0.0046, in good agreement with the perturbative uncertainty of $\pm 0.0047$ 
for the combination of all data.     

\begin{figure}[h]
\begin{center}
\begin{picture}(8.5,9.1)
\put(-1.7,-4.0){\includegraphics{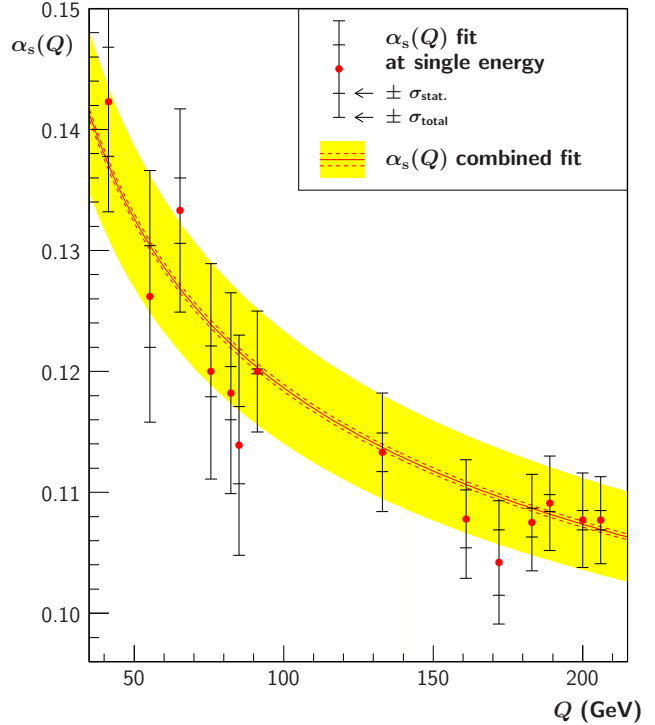}}
\end{picture}
\end{center}
\caption{Energy evolution of $\alpha_s(Q)$.}
\label{fig:running}
\end{figure}

In order to study the energy evolution of $\alpha_s(Q)$ the input measurements from all variables 
and experiments are combined per energy $Q$. The result is shown in fig.~\ref{fig:running}, 
and compared to a the 3-loop evolution fit, which is found in excellent agreement with the 
data ($\chi^2/$N$_{dof}$=11.7/13).  
\section{Conclusions}
\label{sec:conc}
A preliminary combination of measurements of $\alpha_s$ from the LEP experiments using 
event-shape variables is presented. The values of $\alpha_s$ have consistently been determined 
in a common theoretical framework and a new method is applied to assess perturbative 
uncertainties. The combined results are consistent with the expected energy evolution 
of $\alpha_s$. The preliminary result using all LEP data is 
$\alpha_s(M_{\rm Z}) = 0.1201 \pm 0.0048.$  
%

\begin{thebibliography}{}
%
\bibitem{qcdwg} The LEP Experiments ALEPH, DELPHI, L3, OPAL and\\
the LEP QCD Working Group, publication in preparation,\\ 
{\tt http://lepqcd.web.cern.ch/LEPQCD/annihilations/}
\bibitem{inputaleph} The ALEPH Collaboration, CERN-ALEPH 2003-014 CERN-ALEPH CONF 2003-10
\bibitem{inputdelphi} The DELPHI Collaboration, DELPHI 2003-026-CONF-646; DELPHI 2003-019-CONF-639
\bibitem{inputl3} The L3 Collaboration, Phys. Lett. \textbf{B489} (2000) 65;\\
Phys. Lett. \textbf{B536} (2002) 217.
\bibitem{inputopal} The OPAL Collaboration, OPAL Physics Note PN512
\bibitem{theory} R.~W.~L.~Jones, M.~Ford, G.~P.~Salam, H.~Stenzel and D.~Wicke, 
``\,{\it Theoretical uncertainties on $\alpha_s$ from event-shape
variables in $e^+e^-$ annihilations}\,'', in preparation


\end{thebibliography}
%

\end{document}